\documentclass[12pt,twoside]{article}   
\usepackage[super,sort,comma]{natbib}
\usepackage{fancyhdr}		

\usepackage[section]{placeins}   %
\usepackage{graphicx}
\usepackage{amsfonts}
\usepackage{amsmath}

\makeatletter \renewcommand\@biblabel[1]{$^{#1}$} \makeatother
\newcommand{\cen}[1]{\begin{center} #1 \end{center}}

\usepackage{hyperref}
\hypersetup{ colorlinks,
    citecolor=blue,
    filecolor=blue,
    linkcolor=blue,
    urlcolor=blue
}

\usepackage{xcolor}
\definecolor{gray}{rgb}{0.6,0.6,0.6}
\definecolor{red}{rgb}{0.85,0,0}
\definecolor{green}{rgb}{0,0.85,0}
\definecolor{blue}{rgb}{0,0,0.85}
\definecolor{beige}{rgb}{0.92,0.87,0.78}
\usepackage[all]{hypcap}    

\begin{document}

\cen{\sf {\Large {\bfseries A Blind Deconvolution Technique Based on Projection Onto Convex Sets for Magnetic Particle Imaging }} \\  
\vspace*{10mm}


{\Large Onur Yorulmaz $^{*,1}$, Omer Burak Demirel $^{1,2,3,4}$, Yavuz Muslu $^{1,2,5,6}$, Tolga \c{C}ukur $^{1,2,7}$, Emine Ulku Saritas $^{1,2,7}$, Ahmet Enis \c{C}etin $^{1,8}$}\\

\vspace*{3mm}

$^{1}$Department of Electrical and Electronics Engineering, Bilkent University, Ankara, Turkey\\
$^{2}$National Magnetic Resonance Research Center (UMRAM), Bilkent University, Ankara, Turkey\\
$^{3}$Department of Electrical and Computer Engineering, University of Minnesota, Minneapolis, MN, USA\\
$^{4}$Center for Magnetic Resonance Research, University of Minnesota, Minneapolis, MN, United States of America\\
$^{5}$Department of Biomedical Engineering, University of Wisconsin-Madison, Madison, WI, USA\\
$^{6}$Department of Radiology, University of Wisconsin-Madison, Madison, WI, USA\\
$^{7}$Neuroscience Program, Sabuncu Brain Research Center, Bilkent University, Ankara, Turkey\\
$^{8}$Department of Electrical and Computer Engineering, University of Illinois at Chicago, Chicago, IL, USA\\

}

\pagenumbering{roman}
\setcounter{page}{1}
\pagestyle{plain}
(*) Corresponding author, email: yorulmazonur@gmail.com \\

\newpage
\begin{abstract}
\setlength{\baselineskip}{0.7cm} 
\noindent {\bf Purpose:} Magnetic Particle Imaging (MPI) is an emerging imaging modality that maps the spatial distribution of magnetic nanoparticles. The x-space reconstruction in MPI results in highly blurry images, where the resolution depends on both system parameters and nanoparticle type. Previous techniques to counteract this blurring rely on the knowledge of the imaging point spread function (PSF), which may not be available or may require additional measurements. This work proposes a blind deconvolution algorithm for MPI to recover the precise spatial distribution of nanoparticles. 
\\
{\bf Methods:} The proposed algorithm exploits the observation that the imaging  PSF in MPI has zero phase in Fourier domain. Thus, even though the reconstructed images are highly blurred, phase remains unaltered. We leverage this powerful property to iteratively enforce consistency of phase and bounded $\ell_1$ energy information, using an orthogonal Projections Onto Convex Sets (POCS) algorithm. To demonstrate the method, comprehensive simulations were performed without and with nanoparticle relaxation effects, and at various noise levels. In addition, imaging experiments were performed on an in-house MPI scanner using a three-vial phantom that contained different nanoparticle types. Image quality was compared with conventional deconvolution methods, Wiener deconvolution and Lucy-Richardson method, which explicitly rely on the knowledge of PSF. \\
{\bf Results:} Both the simulation results and experimental imaging results show that the proposed blind deconvolution algorithm outperforms the conventional deconvolution methods. Without utilizing the imaging PSF, the proposed algorithm improves image quality and resolution even in the case of different nanoparticle types, while displaying reliable performance against loss of the fundamental harmonic, nanoparticle relaxation effects, and noise. \\
{\bf Conclusions:} In this study, we show the zero-phase property of the imaging PSF in MPI, and exploit this property to reliably recover the nanoparticle distribution without requiring the knowledge of the imaging PSF. By enforcing Fourier transform phase and bounded $\ell_1$ energy constraints, the proposed blind deconvolution algorithm achieves significantly improved results when compared to conventional deconvolution methods.  
\end{abstract}


\newpage     



\setlength{\baselineskip}{0.7cm}      

\pagenumbering{arabic}
\setcounter{page}{1}
\pagestyle{fancy}

\newcommand{\imageLetter}{c}
\newcommand{\ImageLetter}{C}

\section{Introduction}
\label{sec:introduction}
Magnetic particle imaging (MPI) is an emerging biomedical imaging modality \cite{gleich, saritas, goodwill, zheng2017}, with a broad spectrum of applications including angiography \cite{weizenecker2009,lu,vogel2016}, cancer imaging \cite{yu2017}, stem cell tracking \cite{zheng,them2016}, lung perfusion imaging \cite{zhou2017}, temperature mapping \cite{stehning2017}, and viscosity mapping \cite{utkur2017,muslu2018}. MPI images the spatial distribution of superparamagnetic iron oxide (SPIO) nanoparticles by leveraging their nonlinear magnetization response. Since background tissues do not generate any response, they are transparent in MPI. This feature provides an exceptional contrast level, making MPI highly suited for both quantitative and qualitative imaging applications. 

There are two main image reconstruction schemes in MPI: system function reconstruction (SFR) and x-space reconstruction. SFR solves an inverse problem by relying on a lengthy calibration scan that fully characterizes the imaging system \cite{knopp2010,storath2017}. X-space reconstruction, on the other hand, directly grids the speed-compensated MPI signal to the instantaneous position of the scanned field free point (FFP). The main advantage of this technique is that it does not require any calibration scans. Without a calibration scan, however, the resulting images are blurred by the system response and may be sensitive to non-idealities \cite{ilbey2017}.

In x-space image reconstruction, the MPI image is expressed as the convolution of the SPIO distribution with a point spread function (PSF) \cite{goodwill2,goodwill3}. The width of this PSF and thereby the image resolution strongly depend on both the nanoparticle type and the gradient of the selection field that generates the FFP. While the resolution can be improved by using higher selection field gradients, this causes the imaging field-of-view (FOV) to get smaller in return. Hence, there is an intrinsic trade-off between resolution and FOV size in MPI.

Currently, the MPI scanners available commercially or in research settings are designed solely for small-animal imaging. Successful translation to human-sized scanners requires careful evaluation of the human safety limits of magnetic fields in MPI \cite{saritas2, saritas3, schmale2, schmale,demirel}, and these limits render the resolution vs. FOV trade-off more challenging. For human-sized MPI, a typical drive field at or above 25 kHz would be restricted to a maximum amplitude of approximately 7 mT \cite{saritas2,schmale2}. For a typical selection field of 3 T/m/$\mu_0$ and SPIOs with 25~nm diameter, the FOV is limited to 4.7~mm with an image resolution of approximately 1.5~mm \cite{goodwill2,goodwill3}. While millimeter-scale resolution is acceptable in many applications, subcentimeter FOV is quite restrictive for imaging the human body. To overcome this FOV limitation, overlapping patches (i.e., partial FOVs) can be imaged separately and later stitched to increase the overall FOV size \cite{lu}. However, this procedure significantly increases the total imaging time. An alternative strategy is to expand the partial FOV size by lowering the selection field strength, which in turn further worsens the image resolution. To counteract the blurring in x-space MPI reconstructed images, Wiener deconvolution \cite{bente} or spatial-frequency domain equalization techniques \cite{lu2015} have previously been utilized. Both of these approaches rely on the knowledge of the imaging PSF, which may not be available or may require additional measurements.

In this work, we propose a blind deconvolution technique for MPI to recover the precise spatial distribution of nanoparticles. The technique exploits the observation that the imaging PSF in x-space MPI is a zero-phase filter, i.e., the transfer function of MPI is both real-valued \textit{and} positive. Accordingly, the phases of the Fourier transform (FT) coefficients remain the same between the ideal MPI image (i.e., the exact nanoparticle distribution) and the blurry MPI image. We leverage this powerful property as a Fourier-domain phase constraint that is enforced iteratively, using a projection-onto-convex-sets (POCS) algorithm. The proposed algorithm further employs image-domain constraints to limit the spatial support and $\ell_1$ norm of the MPI images. Through successive iterations, it recovers a solution that simultaneously satisfies both the Fourier-domain and image-domain constraints. We present comprehensive simulation results to demonstrate the reliability of the proposed method against a wide range of noise levels, as well as direct feedthough filtering and nanoparticle relaxation effects. With imaging experiment results on our in-house FFP MPI scanner, we validate that the proposed method achieves significantly improved results in comparison to conventional deconvolution techniques that require the knowledge of the imaging PSF.

\section{THEORY AND METHODS}
\label{theory}

The proposed blind deconvolution technique employs an iterative POCS algorithm that deblurs an MPI image based on several constraints in Fourier and image domains. These constraints are concerned with the phase of the Fourier domain coefficients, spatial support of the image, and spatial regularization of the image. In this work, we exploit the observation that the imaging PSF in MPI is a zero-phase filter in Fourier domain, causing the ideal and blurred images to have matched phases. In this section, we start by explaining the rationale for the zero-phase property. We then present the proposed method by describing the individual Fourier- and image-domain constraints, and their implementation in the POCS algorithm. Finally, we describe the simulations and imaging experiments that are used to demonstrate our method. 

\subsection{Zero-Phase Property of the Imaging PSF}
According to the x-space MPI theory \cite{goodwill2,goodwill3}, MPI images can be formulated as a convolution of the nanoparticle distribution with the imaging PSF, i.e.,
\begin{equation}
    g(x,y,z) = \imageLetter_o*h(x,y,z)
\end{equation}
where $\imageLetter_o : \mathbb{R}^3 \rightarrow \mathbb{R}_+$ is the spatial distribution of magnetic particle concentration, $h : \mathbb{R}^3 \rightarrow \mathbb{R}_+$ is the 3D PSF, and $g : \mathbb{R}^3 \rightarrow \mathbb{R}_+$ is the PSF-blurred MPI image.

\begin{figure}[ht!]
\centering
\includegraphics[width=0.8\columnwidth]{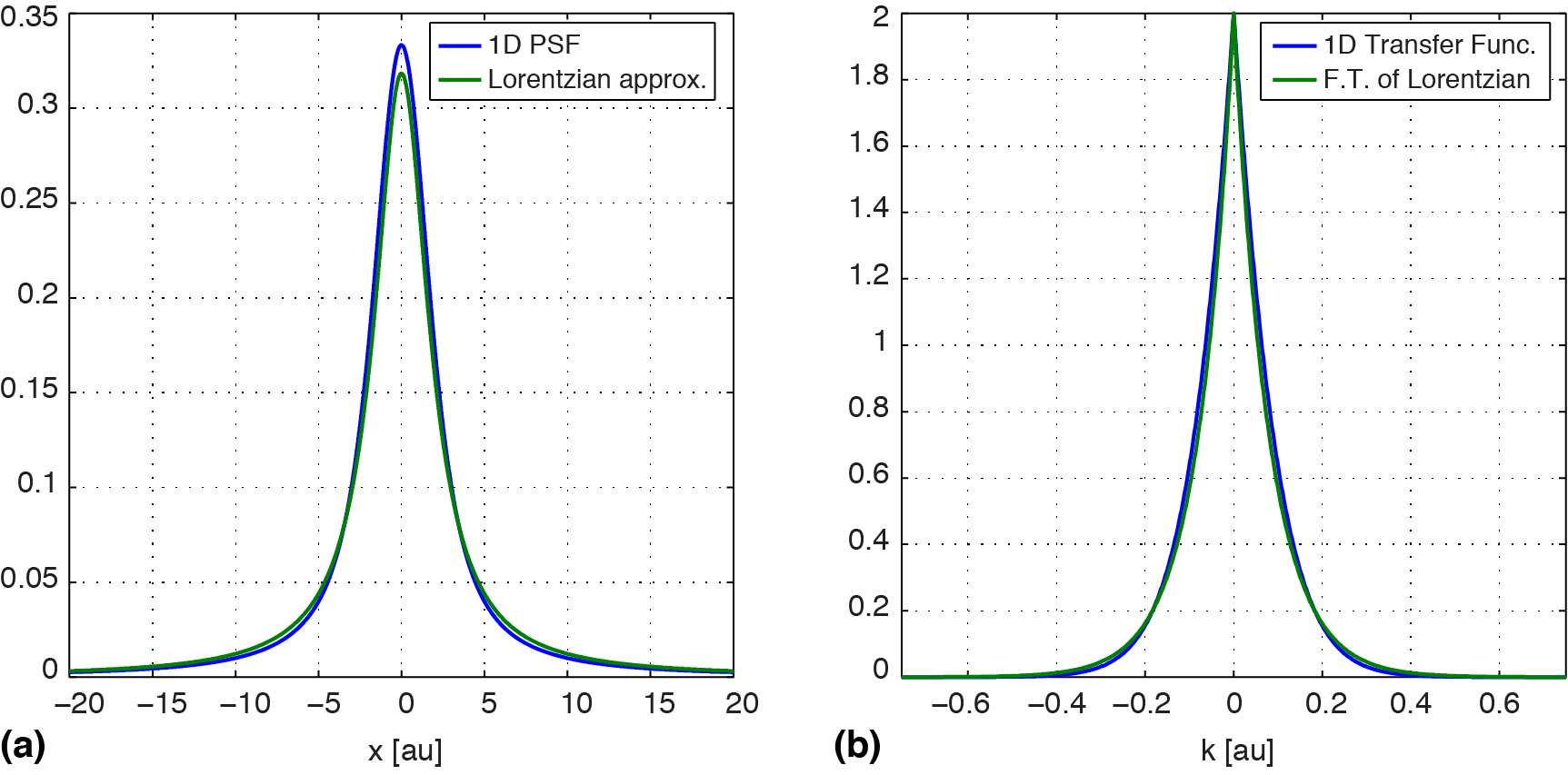}
\caption{(a) One-dimensional (1D) PSF of MPI (i.e., the derivative of the Langevin function) and its Lorentzian approximation. (b) The corresponding 1D transfer function and the Fourier Transform of the Lorentzian. The 1D transfer function is a purely real-valued and zero-phase function.\newline}
\label{1dRealValued}
\end{figure}

Starting with the simple case of one-dimensional (1D) imaging, the PSF can be expressed as:
\begin{equation}
    h(x) = \mathcal{L}'(G x / H_{sat}).
\end{equation}
Here, $H_{sat}$ [T/$\mu_0$] is the magnetic field required to half-saturate the nanoparticles, and $G$ [T/m/$\mu_0$] is the gradient of the selection field. In addition, $\mathcal{L}'$ denotes the derivative of the Langevin function, $\mathcal{L}$, which characterizes the nonlinear magnetization response of the nanoparticles. $\mathcal{L}'$ can be written as follows:
\begin{equation}
    \mathcal{L}'(x) = \frac{1}{x^2} - \frac{1}{sinh^2(x)}.
\end{equation}
Although its Fourier transform does not have a closed form expression, $\mathcal{L}'$ can be well approximated (within $2\%$ error) by the following Lorentzian function \cite{goodwill2}:
\begin{equation}
    \tilde {\mathcal{L}}(x) = \frac{2}{\pi} \frac{2}{4+x^2}.
\end{equation}
This Lorentzian function has a closed-form Fourier transform, which can be expressed as:
\begin{equation}
    \mathcal{F}\{\tilde {\mathcal{L}}(x)\}(k) = 2e^{-4\pi|k|}
\end{equation}
As seen in this expression, the transfer function based on the Lorentzian approximation is a zero-phase function, i.e., it is both \textit{real-valued} and \textit{positive} at all \textit{k}-values. Similarly, the exact PSF, based on smooth and differentiable Langevin function, also has a zero-phase Fourier transform as demonstrated by numeric calculations shown in Figure \ref{1dRealValued}. 

\begin{figure}[ht!]
\centering
\includegraphics[width=0.8\columnwidth]{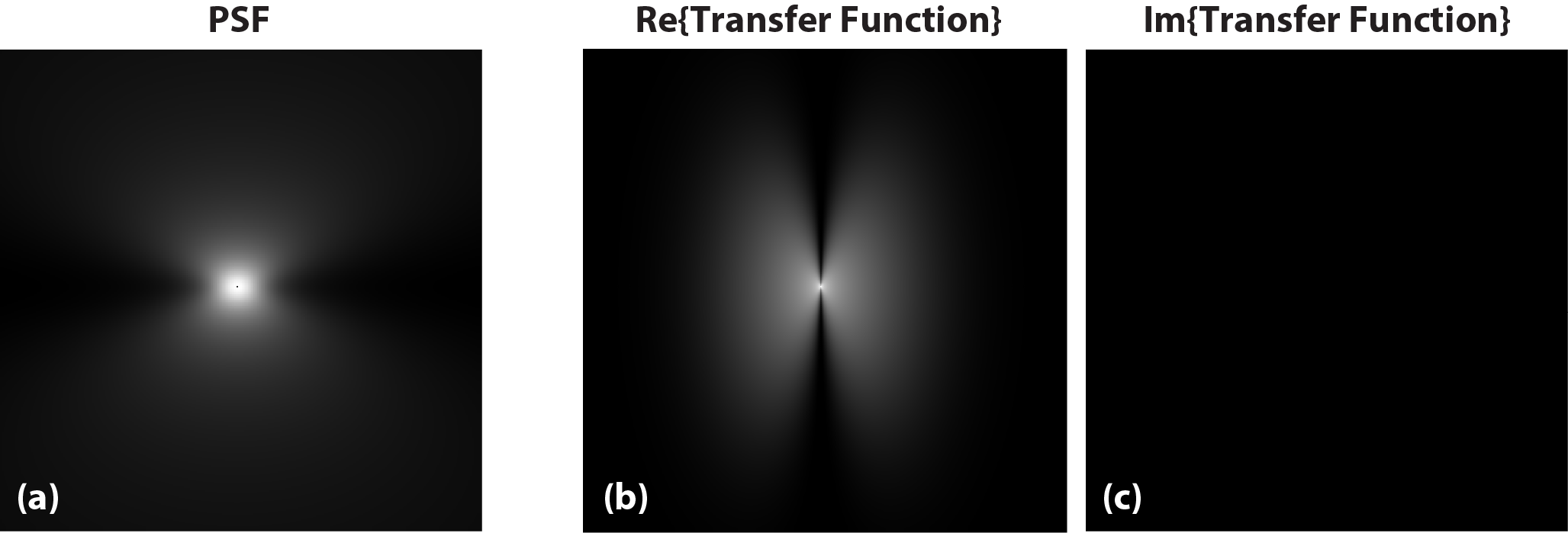}
\caption{(a) Two-dimensional (2D) point spread function (PSF) of MPI. The corresponding transfer function is a zero-phase function. (b) The real part of the transfer function is positive-valued, and (c) the imaginary part is completely zero (here (b) and (c) are displayed with identical windowing). \newline}

\label{2dRealValued}
\end{figure}

In multi-dimensional case, while the closed-form expressions for the PSF have been derived \cite{goodwill3}, there is no closed form expression for the transfer function. To assess the phase of the transfer function, here we performed a numerical evaluation of the PSF and its Fourier transform, with 2D results displayed in Figure \ref{2dRealValued}. While the numerical evaluations were performed only at a finite number of discrete frequencies,  
it can be demonstrated that the zero-phase feature also applies in 2D and 3D, i.e.,
\begin{equation}
    \angle{\mathcal{F}\{h(\textbf{x})\}(\textbf{k})} = 0 \:\:\: \forall \:\textbf{k}
\end{equation}
or,
\begin{gather}
    Re\{\mathcal{F}\{h(\textbf{x})\}(\textbf{k})\} \geq 0  \:\:\: \forall \:\textbf{k} \\
    Im\{\mathcal{F}\{h(\textbf{x})\}(\textbf{k})\} = 0  \:\:\: \forall \:\textbf{k}
\end{gather}
where $Re$ and $Im$ represents the real and imaginary parts, and the operator $\angle$ represents the angle of the complex number.

Note that while any symmetric PSF would have a real-valued transfer function, the zero-phase property is a much stronger and less common feature. Here, we leverage this powerful property of the PSF in deblurring of MPI images, as explained in the next section.

\subsection{Fourier-Domain Phase Constraint}
\label{Fourier}
In this section, we describe a closed and convex set based on the phase of the Fourier transform coefficients of MPI images. Let $\imageLetter_o [n_1,n_2] : \mathbb{Z}^2 \rightarrow \mathbb{R}_+$ be the 2D ideal MPI image (i.e., the exact spatial distribution of nanoparticle concentration) and $h[n_1,n_2] : \mathbb{Z}^2 \rightarrow \mathbb{R}_+$ be the 2D PSF, sampled at discrete spatial locations $0\leq n_1\leq N_1-1$ and $0\leq n_2\leq N_2-1$ along the x- and y-axis, respectively.  The blurry MPI image $g[n_1,n_2] : \mathbb{Z}^2 \rightarrow \mathbb{R}_+$ can be modelled as the convolution of $\imageLetter_o$ with $h$:

\begin{equation}
    g[n_1,n_2] = \imageLetter_o * h[n_1,n_2],
\label{convolution}
\end{equation}
where ``$*$" represents the two-dimensional convolution operator. The Discrete-Time Fourier transform (DTFT) of $g$ is, therefore, given by:
\begin{equation}
    G(\omega_1, \omega_2) = \ImageLetter_o (\omega_1, \omega_2) \cdot H(\omega_1, \omega_2),
\end{equation}
where $\ImageLetter_o$ and $H$ are the two dimensional DTFT of $\imageLetter_o$ and $h$, respectively, and ``$\cdot$" denotes element-wise multiplication. With the assumption that $H(\omega_1, \omega_2)$ is real and positive, the phase of $\imageLetter_o$ can be exactly recovered from the phase of $G$.


The above-mentioned phase recovery can be implemented exactly for any PSF that satisfies the zero-phase condition. Furthermore, it can be extended to cases where the PSF is symmetric and thus the transfer function is real (but not necessarily always positive). In such cases, we can either estimate the locations of phase jumps of $\pi$ in $G$ or else perform phase unwrapping. Afterwards, the phase of $\ImageLetter_o$ can be accurately recovered from the phase of $G$.


Here, we assume that the following closed and convex set \cite{youla} based on phase information is available in MPI:
\begin{equation}
    S_\phi = \left\{\imageLetter \  |  \ \angle \ImageLetter(\omega_1, \omega_2)= \Phi (\omega_1, \omega_2) \right\}
    \label{c_theta}
\end{equation}
Here, $S_\phi$ is the set of images that possess a given Fourier transform phase. This transform phase $\Phi$ is estimated from the phase of the blurry image $\angle{\mathcal F\{g\}}$. An orthogonal projection $\imageLetter_p$ of an MPI image $\imageLetter$ onto this set will enforce the desired phase property,
\begin{equation}
    \ImageLetter_p (\omega_1, \omega_2)=  |\ImageLetter (\omega_1, \omega_2)|. e ^{j  \Phi (\omega_1, \omega_2)} 
    \label{Cp_theta}
\end{equation}
where $\ImageLetter_p$ is the DTFT of the projection image $\imageLetter_p$ and $|\ImageLetter (\omega_1, \omega_2)|$ is the magnitude of the DTFT of $\imageLetter$. The projection operation simply replaces the phase of $\ImageLetter$ with $\Phi (\omega_1, \omega_2)$.

\subsection{Spatial-Domain Support Constraint}
\label{Spatial}
The phase information in Eq.~\ref{Cp_theta} is not sufficient to determine a solution to the inverse problem. 
One powerful solution is to constrain the spatial-domain support of the ideal image $\imageLetter_o$, as proposed in magnitude retrieval algorithms studied in 1980s \cite{oppenheim, youla, sezan, yorulmaz1}. Here, we assume that the following closed and convex set based on support information is available,
\begin{equation}
    S_s = \left\{\imageLetter \ | \ \imageLetter[n_1,n_2] = 0,  \  \ [n_1,n_2]  \not \in \ S
    \right\}
\end{equation}
where $S$ is the spatial support of the ideal image $\imageLetter_o$.

The  projection $\imageLetter_s$ of an arbitrary image $\imageLetter$ onto the spatial-domain constraint set $S_s$ is performed by simply retaining the portion of the image within the support region:
\begin{equation}
    {\imageLetter_s}[n_1,n_2] =  \left\{
                \begin{array}{ll}
                  \imageLetter[n_1,n_2], &if \ \ [n_1,n_2] \in S\\
                  0, &if \  \ [n_1,n_2] \not \in S\\
                \end{array}
              \right.
\label{supportProj}
\end{equation}

This step and the previous phase-constraint step are performed in an iterative manner, successively imposing the sets $S_\phi$ and $S_S$ on the iterates \cite{oppenheim, youla, sezan, yorulmaz1}. In practice, the precise support region $S$ of the ideal MPI image may not be available. A reasonable choice for $S$ would then be the imaging FOV (i.e., the region covered by the FFP trajectory), as done in this work. For that case, the spatial-domain constraint is equivalent to restricting the spatial support of the nanoparticle distribution to the FOV. 


\subsection{Spatial-Domain Regularization Constraint}
\label{L1balls}
Noise in MPI acquisitions can degrade the accuracy of the estimates for both the Fourier-domain phase and the spatial support, thereby compromising deconvolution performance. To improve performance, here we employ spatial regularization in image domain by defining a close and convex set $S_1^\epsilon$ based on $\ell_1$ norm \cite{combettes},
\begin{equation}
     S_1^\epsilon = \left\{\imageLetter \ | \ ||\imageLetter||_1  \leq \epsilon \right\}
\end{equation}
where $\epsilon$ is a predetermined upper bound, and
\begin{equation}
    ||\imageLetter||_1 = \sum_{n_1,n_2} |\imageLetter[n_1,n_2]|.
\end{equation}
This regularization can be implemented through an orthogonal projection onto $S_1^\epsilon$ \cite{cetin, tofighi1,brucker}. It is also possible to select the upper bound $\epsilon$ in an adaptive manner using the epigraph set concept described in \cite{yorulmaz1}. 

\begin{figure}[ht!]
\centering
\includegraphics[width=0.5\columnwidth]{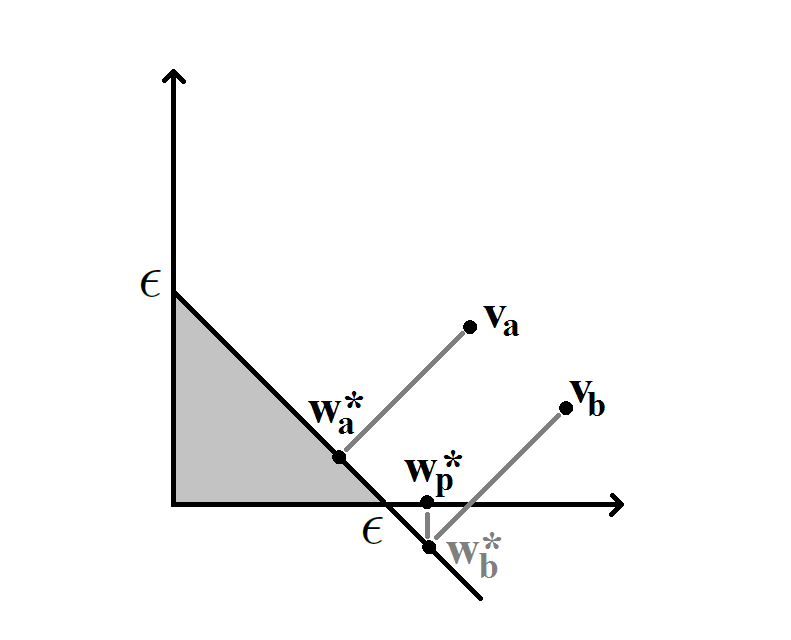}
\caption{Shaded area represents the first quadrant of the 2D version of $\ell_1$ ball. The vector $\mathbf{w_a^*}$ is the projection of $\mathbf{v_a}$ onto the $\ell_1$ ball. The projection of $\mathbf{v_b}$ is difficult to compute in high dimensions, but $\mathbf{w_p^*}$ can be used instead of the projection.\newline}
\label{fig:firstQuad}
\end{figure}

In this paper, we use an approximate projection operator to speed up the iteration process. The magnitudes of the result of the projection is independent of the signs, therefore it is possible to consider the absolute values of the pixels and incorporate the original signs later. As a result, we consider only the first `quadrant' of the $\ell_1$ ball as shown in Figure \ref{fig:firstQuad}.

Let $\mathbf{v_a}$ be the vectorized version of the current iterate $\imageLetter^{(k)}$ image, i.e., $\mathbf{v_a}[i] = \imageLetter^{(k)}[n_1,n_2]$ for $i=n_1+n_2\cdot N_1$. In Figure \ref{fig:firstQuad}, $\mathbf{w_a^*}$ is the orthogonal projection of $\mathbf{v_a}$ onto the $\ell_1$ ball. The vector $\mathbf{w_a^*}$ is the solution of the following problem:
\begin{equation}
    \mathbf{w_a^*} = \text{argmin} ||\mathbf{v_a} - \mathbf{w}||_2^2 \ such\ that \  \sum_{i = 0}^{N-1} |w[i]| \leq \epsilon
\end{equation}
where $N=N_1\cdot N_2$ is the total number of pixels in the image $\imageLetter^{(k)}$. Since we only consider the all-positive part of the $\ell_1$ ball, the solution can be found by minimizing the Lagrangian.
\begin{equation}
    L = \frac{1}{2} ||\mathbf{w} - \mathbf{v_a}||^2 + \lambda (\sum_{i = 0}^{N-1} w[i] - \epsilon)
\end{equation}
where $\sum_{i = 0}^{N-1} w[i] = \epsilon$ is the hyperplane determining the boundary of the $\ell_1$ ball.

By computing the partial derivatives with respect to the entries of unknown vector $\mathbf{w}$ and the Lagrange multiplier $\lambda$, we can determine the projection of $\mathbf{v_a}$ onto the hyperplane $\sum_{i = 0}^{N-1} w[i] = \epsilon$ as follows:
\begin{equation}
    w[i] = v_{a}[i] - \lambda
\end{equation}
and 
\begin{equation}
    \lambda = \frac{1}{N}\left( \sum_{i = 0}^{N-1} v_{a}[i] - \epsilon \right)
\end{equation}
Therefore, the vector $\mathbf{w_a^*}$ is given by
\begin{equation}
    w_{a}^*[i] = v_{a}[i] - \frac{1}{N} \left(\sum_{i=0}^{N-1} v_{a}[i] - \epsilon \right)
\end{equation}

The vector $\mathbf{w_a^*}$ may or may not be the orthogonal projection onto the $\ell_1$ ball, as shown in Figure \ref{fig:firstQuad}. If all the entries $w_{a,i}$ of $\mathbf{w_a^*}$ are positive, it is the projection vector. However, some of the entries of the projection vector $\mathbf{w_b^*}$ corresponding to $\mathbf{v_b}$ may become negative, as shown in Figure \ref{fig:firstQuad}. In this case we zero out the negative entries of $\mathbf{w_b^*}$ and obtain $\mathbf{w_p^*}$ as shown in Figure \ref{fig:firstQuad}:
\begin{equation}
    w_{p}^*[i] = \left\{
                \begin{array}{ll}
                  w_{b}^*[i]  , &\ if\ w_{b}^*[i] \geq 0\\
                  0 , &\ if\  w_{b}^*[i] < 0\\
                \end{array}
              \right.
              \label{l1baba}
\end{equation}
Although the vector $\mathbf{w_p^*}$ can be further processed to determine the projection vector onto the $\ell_1$ ball, we use $\mathbf{w_p^*}$ in our algorithm. This speeds up the iteration process, because $\ell_1$ projection requires sorting all the elements of a given image \cite{duchi}. By using $\mathbf{w_p^*}$, we eliminate the sorting process.

Equation (\ref{l1baba}) is very useful during iterations, because it not only removes noise but also suppresses small image coefficients, enforcing the image energy to be concentrated on stronger coefficients.

\subsection{Proposed Blind Deconvolution Algorithm}
\label{deconvalgorithm}
Three closed and convex sets for Fourier domain phase, spatial-domain support, and spatial regularization were defined in Sections \ref{Fourier} through \ref{L1balls}. The proposed deconvolution algorithm employs iterative projections onto these sets, $S_\phi$, $S_s$, and $S_1^\epsilon$ to simultaneously enforce the respective image properties. Let $\imageLetter^{(k)}$ be the current, and $\imageLetter^{(k+1)}$ be the next iterate for the deconvolved MPI image. The projections are implemented during the iterative deconvolution as follows: 
\begin{equation}
    \imageLetter^{(k+1)} = P_1 P_s P_\phi (\imageLetter^{(k)})
    \label{iterationAlgorithm}
\end{equation}
Here, $P_1$ is the projection operator onto the $\ell_1$ ball (Section \ref{L1balls}), $P_s$ is the projection operator onto the MPI spatial support set (Section \ref{Spatial}), and $P_\phi$ is the projection operator onto the phase set (Section \ref{Fourier}). Note that these projection operators do not rely on estimates of the imaging PSF at any stage. Since the proposed algorithm does not utilize PSF information, it is a blind deconvolution method.

The POCS theorem \cite{youla} states that successive projections onto multiple sets will converge to a solution at the intersection of the sets, assuming that the intersection is non-empty. Thus iterates $\imageLetter^{(k)}$ will converge to an image in $S_1 \cap S_s \cap S_\phi $ regardless of the initial image $\imageLetter^{(0)}$. When the convex sets do not intersect, the iterates oscillate between three images satisfying either one of the sets $S_1$, $S_s$, and $S_\phi$. At that point, the iterations and one of the above-mentioned solutions can be accepted as the solution to the inverse problem \cite{Combettes1,Combettes2,Trussell1,Trussell2}. In practice, iterations can be stopped when the difference between consecutive iterates become insignificant. In this work, it was observed that a total of 100 iterations were sufficient to obtain stable results.

In the remainder of this section, we give implementation details for the algorithm outlined in Equation \ref{iterationAlgorithm}. We first start by projecting the current image onto convex set $S_\phi$ defined by Equation \ref{c_theta}. Since the PSF has zero-phase characteristics, we assume that the ideal and blurry MPI images carry identical phase information. However, this assumption is typically violated due to the finite FOV imaged, which causes the observed MPI image to be a truncated version of the full convolution of the ideal image and the PSF. This truncation forces a narrow rectangular support in spatial domain, which corresponds to a convolution in the Fourier domain that can disturb both the phase and magnitude of DTFT coefficients. Therefore, depending on the severity of the truncation, the phase information regarding the ideal image is only approximate.

As we have shown previously, Fourier transform phase can be recovered using a gradient-based algorithm in order to deconvolve microscopic images in a blind manner \cite{yorulmaz1}. To address the truncation problem here, an estimate of non-cropped blurry image is obtained by replication of the boundaries \cite{hamey2015}. For this purpose, we first stretch the boundary pixels of the blurry image in all directions as follows:

\begin{equation}
\resizebox{.89\hsize}{!}{
    $\hat{g}[n_1,n_2]=\left\{
                \begin{array}{ll}
      g[n_1-N_e,n_2-N_e],&N_e \le n_1 \le N_e + N_1, \:\: N_e \le n_2 \le N_e + N_2\\
      g[n_1-N_e,0],&N_e \le n_1 \le N_e + N_1, \:\: n_2 < N_e\\
      g[n_1-N_e,N_2-1],&N_e \le n_1 \le N_e + N_1, \:\:n_2 > N_2+N_e\\
      g[0,n_2-N_e],&N_e \le n_2 \le N_e + N_2, \:\:n_1 < N_e\\
      g[N_1-1,n_2-N_e],&N_e \le n_2 \le N_e + N_2, \:\:n_1 > N_1+N_e\\
      g[0,0],&n_1 < N_e, \:\:n_2 < N_e\\
      g[N_1-1,N_2-1],&n_1> N_1+N_e, \:\:n_2 > N_2+N_e\\
      g[0,N_2-1],&n_1 < N_e, \:\:n_2 > N_2+N_e\\
      g[N_1-1,0],&n_2 < N_e, \:\:n_1 > N_1+N_e\\
                \end{array}
              \right.
              $}
              \label{imextend}
\end{equation}

Here, $\hat{g}$ is the stretched image and $N_e$ is the size of the extension. The size of the stretched image $\hat{g}$ then becomes $2N_e + N_1$ by $2N_e + N_2$. 

\begin{figure}
\centering
\includegraphics[width = 0.7\columnwidth]{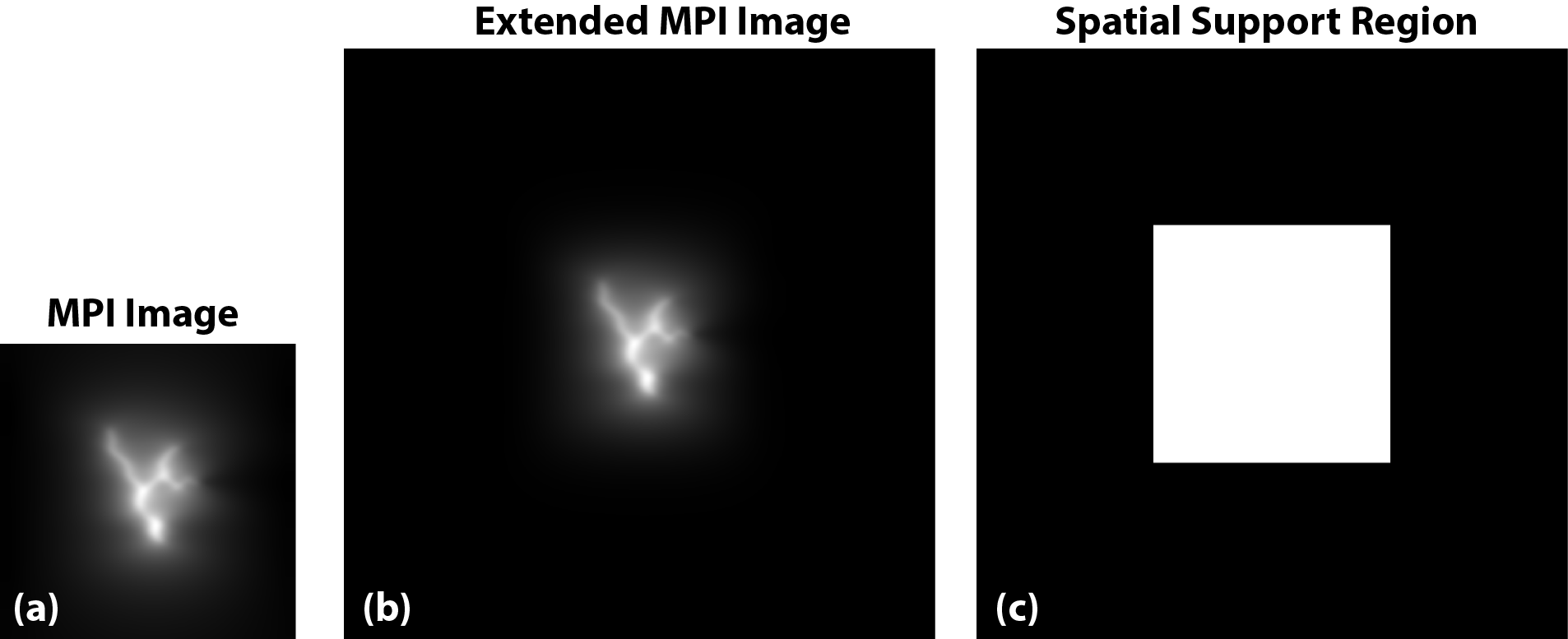}
\caption{(a) The blurry MPI image, (b) its extended version, and (c) the support that marks the nonzero region used in spatial domain projections. Here, the spatial support is the same size as the imaging FOV.\newline}
\label{supportfig}
\end{figure}

Note that a pure stretching of the pixels at the image boundaries will not reflect the PSF-convolved output of the ideal image over an extended FOV. To closely mimic the effects of convolution via a natural PSF, the pixel intensities need to gradually fade to zero. To obtain such a gradual fading effect, we apply a gradient mask to the stretched image. This gradient mask is obtained by convolving the spatial support, $s$, of the blurry image with a Gaussian low-pass filter, $h_G$, with large variance (in this work, $\sigma$ was chosen as $10\%$ of $N_1$ and $N_2$ in both directions). The final extended image, $\tilde{g}$, is then obtained via multiplying the stretched image with the gradient mask, i.e.,
\begin{equation}
\tilde{g} = \hat{g} \cdot (s * h_G),
\label{windowing}
\end{equation}
where, 
\begin{equation}
\resizebox{.7\hsize}{!}{
  $  s[n_1,n_2] = \left\{
                \begin{array}{ll}
                  1,& N_e \le n_1 \le N_e + N_1, \:\: N_e \le n_2 \le N_e + N_2\\
                  0,& o.w.\\
                \end{array}
              \right.$
}
\end{equation}

These steps ensure that the extended image matches the blurry image in central locations, while gradually fading to zero in the extended regions. In Figure \ref{supportfig}, an example blurry image, its extended version, and the spatial support region are shown. For all steps of the proposed algorithm, the extended image is utilized. During implementation, the $\epsilon$ parameter of the proposed $l_1$ ball projection in each iteration was set to be the 98\% of the $l_1$ norm of the iterate.




\subsection{Comparison with Conventional Deconvolution Methods}
For comparison, we also performed deconvolution on blurry MPI images with the conventional Wiener and Lucy-Richardson methods, using the respective algorithms available in MATLAB (Mathworks, Natick, MA, USA). Note that neither of these algorithms is blind, but rather explicitly require information about the imaging PSF. Here, we utilized the exact PSF for both of these algorithms in the case of simulations. For the results from the imaging experiment, we utilized the same PSF as in simulations. The noise-to-signal power ratio parameter of Wiener deconvolution was set to 1.0. The iteration number parameter of Lucy-Richardson was set to 100. The comparison of techniques was performed using the peak signal-to-noise ratio (PSNR) metric, which is defined as follows:

\begin{equation}
    PSNR = 10  \text{log}_{10} \left ( \frac{MAX^2}{MSE} \right ) 
\end{equation}
Here, $MAX$ is the peak signal intensity of the ideal image and $MSE$ is the root-mean-squared error between the ideal and the deconvolved image. PSNR is evaluated in dB, with higher values corresponding to improved image quality. 

\subsection{Simulations}
MPI simulations were performed on a custom toolbox developed in MATLAB, using the multi-dimensional x-space formalism \cite{goodwill3}. The selection field had gradients $(-6,3,3)$ $T/m/\mu_0$ in x-,y-, and z-directions, respectively, comparable to that of a previously reported FFP MPI scanner \cite{lu}. The nanoparticle diameter was assumed to be 25 nm, a reasonable value given the developments on tailored MPI nanoparticles \cite{ferguson}. A rectilinear trajectory was utilized with an overall FOV of 4 cm $\times$ 4 cm in x-z plane, covered line-by-line by a total of 201 lines in x-direction. The drive field in z-direction was at $f_0$ = 25 kHz with 10 mT amplitude, covering a partial FOV (pFOV) size of 6.6 mm. Neighboring pFOVs had $80\%$ overlap along the z-direction, to allow the recovery of the DC component that is lost due to direct feedthrough filtering \cite{lu}. A single channel receive coil along the z-direction was assumed, and the MPI signal was sampled at a rate of $f_s$ = 25 MS/s.

In MPI, nanoparticle relaxation effects delay and blur the time-domain MPI signal. To test the robustness of the proposed method, these effects were incorporated via an exponential kernel that convolves the adiabatic MPI signal \cite{croft, croft2}: 

\begin{equation}
    s(t) = s_{adiab}*r(t)
    \label{relaxMPI}
\end{equation}
where
\begin{equation}
    r(t) = \frac{1}{\tau}e^{-t/\tau} u(t)
\end{equation}
Here, $s(t)$ is the signal with relaxation effects, $s_{adiab}(t)$ is the adiabatic MPI signal without relaxation delays, and $u(t)$ is the Heaviside step function. This simple but powerful model was previously shown to provide a very good match to x-space MPI data across a wide range of drive field frequencies and amplitudes \cite{croft, croft2}. At the chosen operating frequency and amplitude, a realistic relaxation time constant of $\tau= 0.5$ $\mu s$ was incorporated based on previous experimental results \cite{croft}. After direct feedthrough filtering, the resulting MPI signal for each pFOV was processed via x-space reconstruction \cite{goodwill2,goodwill3,lu} that incorporated a $\tau/2$ time-delayed speed compensation \cite{croft}: 
\begin{equation}
\label{speedcomptau}
    g(x_s(t)) = \frac{s(t)}{\gamma \dot{x}_s(t-\tau/2)}
\end{equation}
Here, $x_s(t)$ is the instantaneous position of the FFP and $\gamma$ is a scalar. In this equation, the time delay in speed compensation widely makes up for the effective position shift in the image that would otherwise be induced by relaxation \cite{croft}. 

For this work, three sets of data were generated:
\begin{enumerate}
  \item PSF-Blurred MPI Image: This image is simply the convolution of the nanoparticle distribution (i.e., phantom) with the 3-dimensional MPI PSF \cite{goodwill3}. 
  \item Reconstructed MPI Images, without Relaxation: These images simulate the signal acquisition process, taking into account the fact that the fundamental harmonic is filtered out (e.g., using a notch filter centered around the drive field frequency) to avoid the direct feedthrough signal. Each pFOV is then reconstructed via x-space MPI by first dividing the filtered MPI signal by the scanning speed (i.e., by setting $\tau$ = 0 in Eq.~\ref{speedcomptau}). The resulting pFOV image was then gridded to the instantaneous position of the FFP \cite{goodwill2}. A subsequent DC recovery algorithm was implemented using the smoothness of the overall MPI image \cite{lu}, and the DC-recovered pFOV images were merged in an SNR-optimized fashion \cite{bozkurt}. To test the SNR robustness of the proposed algorithm, this entire procedure was performed at 9 different SNR levels, by adding additive white Gaussian noise to the simulated MPI signal before the image reconstruction step.
  \item Reconstructed MPI Images, with Relaxation: These images also simulate the entire signal acquisition process, with the addition of relaxation effects using Eq. ~\ref{relaxMPI}. Again, additive white Gaussian noise was added to the MPI signal at 9 different SNR levels, followed by direct feedthrough filtering. The images were reconstructed via x-space reconstruction, using the same procedure as described above.
\end{enumerate}

For the noisy image sets, SNR was defined as the ratio of the peak signal that resulted during the scanning of the entire FOV, divided by the standard deviation of the added white Gaussian noise. We tested 8 different SNR levels between 5 dB and 40 dB, with steps of 5 dB. In addition, a zero-noise case (i.e., infinite SNR) was also tested. 

The imaging phantoms with $4cm\times4cm$ size were sampled on a $4001\times4001$-pixel grid. The imaging PSFs were sampled on a $8001\times8001$-pixel grid, extending over a broad area when compared to the object itself. The full convolution output of the phantom and the imaging PSF was truncated to the original $4001\times4001$-pixel grid of the phantom to generate the PSF-blurred MPI image. Each scanned pFOV had $f_s/2f_0$ samples before gridding (i.e., 500 samples over a pFOV of 6.6 mm for the above-mentioned parameters). After x-space reconstruction, each line was interpolated down to a 1D grid of size $4001$, resulting in a $201\times4001$ image. This image was then fed to the proposed algorithm and the comparison deconvolution methods. Finally, all resulting images were interpolated to $4001\times4001$ for display purposes, as well as quantitative image quality analysis via PSNR. 


\subsection{Imaging Experiment}
For imaging experiment, a phantom containing three 2-mm inner diameter vials was prepared. The vials were placed at 23-mm center-to-center separations along the z-direction. The first vial was filled with Nanomag-MIP nanoparticles (Micromod GmbH, Germany) with a concentration of 1.43 mg Fe/mL. The third vial was filled with Vivotrax nanoparticles (Magnetic Insight Inc., USA) with a concentration of 5.5 mg Fe/mL. The second vial was filled with a homogeneous mixture of these two nanoparticles. 

This phantom was utilized in an imaging experiment performed on our in-house FFP MPI scanner with selection field gradients of $(-4.8,2.4,2.4)$ $T/m/\mu_0$ in x-,y-, and z-directions, respectively \cite{utkurBIYOMUT}. The overall FOV was 0.8 cm $\times$ 11 cm in x-z plane, covered by a total of 9 lines in x-direction. The drive field along the z-direction was applied at $f_0$~=~9.7~kHz with 15 mT amplitude. Neighboring pFOVs had 85\% overlap along the z-direction. A receive coil along the z-direction was utilized, and the MPI signal was sampled at 2 MS/s. The fundamental harmonic was filtered out to remove direct feedthrough effects. The rest of the signal processing and image recontruction steps were identical to those in simulations. 

\begin{figure}[]
\centering
\includegraphics[width = 0.8\columnwidth]{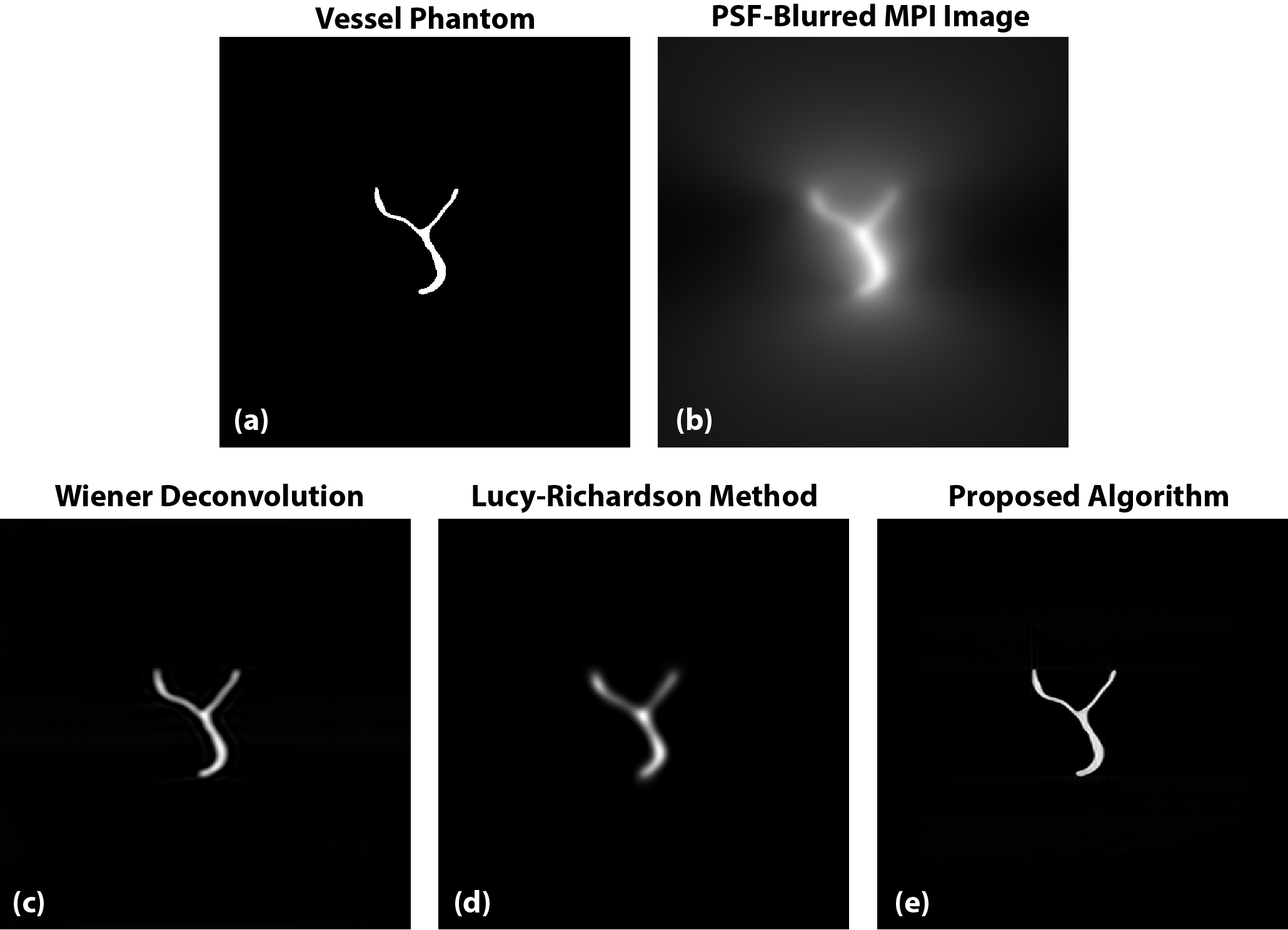}
\caption{(a) The original vein phantom, (b) the PSF-blurred MPI image assuming an SPIO size of 25nm, but no acquisition noise (PSNR = 14.66 dB with respect to the vein phantom). The results of (c) Wiener deconvolution (PSNR = 28.38 dB), (d) Lucy-Richardson method (PSNR = 26.14 dB), and (e) the proposed algorithm (PSNR = 32.26dB) after 100 iterations.\newline
}
\label{idealcase}
\end{figure}

\begin{figure}[t]
\centering
\includegraphics[width = 0.5\columnwidth]{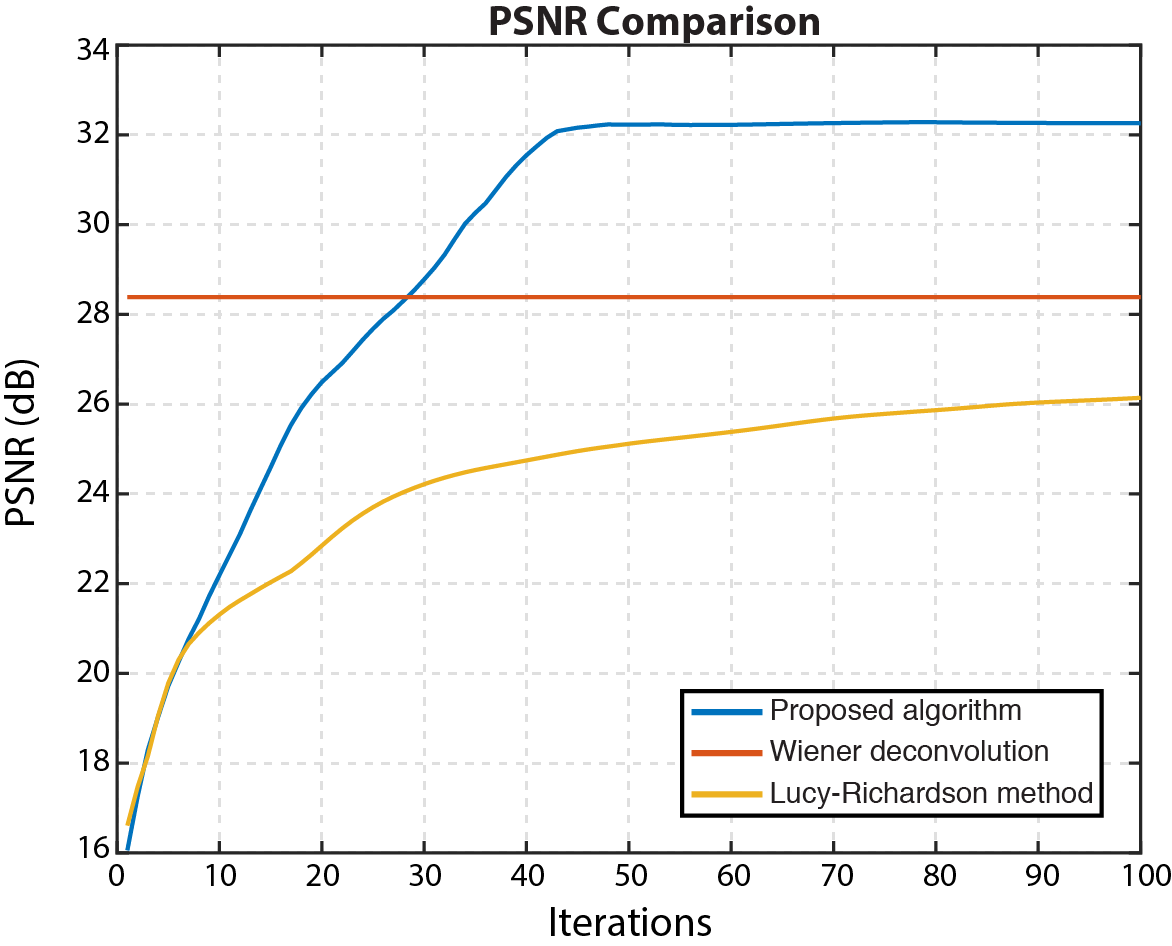}
\caption{PSNR as a function of iteration count, measured between the vein phantom (shown in Figure \ref{idealcase}a) and the deconvolved images for the Lucy-Richardson method and the proposed algorithm. Wiener deconvolution is displayed at a constant value, as it is a one-step algorithm. In this case, the proposed algorithm converges to a stable solution within 40 iterations, at significantly higher PSNR levels when compared to the other two deconvolution methods.\newline}
\label{PSNRperIter}
\end{figure}


\section{RESULTS}

\label{results}
Here, we demonstrate the proposed deconvolution algorithm on MPI images with significant blurring and relaxation effects. In our simulations, we first processed a simple vein phantom to examine the progression of image recovery across iterations for the case of a PSF-blurred MPI image.
The vein phantom (i.e., the ideal image), the PSF-blurred MPI image, and the deconvolution results are shown in Figure \ref{idealcase}. The Wiener deconvolution produces better edge emphasis than the Lucy-Richardson method, but it suffers from ringing artifacts due to resolution loss. Overall, even though the Wiener and Lucy-Richardson methods are PSF-informed, they yield images with significant residual blurring and relatively low PSNR levels. This is despite the fact that a relatively large FOV when compared to the vessel phantom was utilized, which typically improves PSF-informed deconvolution results. In comparison, the proposed blind deconvolution method recovers a deblurred image with higher spatial resolution and PSNR, and one that visually matches the phantom image. The proposed method achieves 3.8 dB and 6.1 dB higher PSNRs than Wiener deconvolution and Lucy-Richardson method, respectively. The convergence speed and PSNR performance per iteration cycle are also given in Figure \ref{PSNRperIter}, showing that the proposed algorithm converges to a stable result in about 40 iterations. 

\begin{figure}[t]
\centering
\includegraphics[width = \columnwidth]{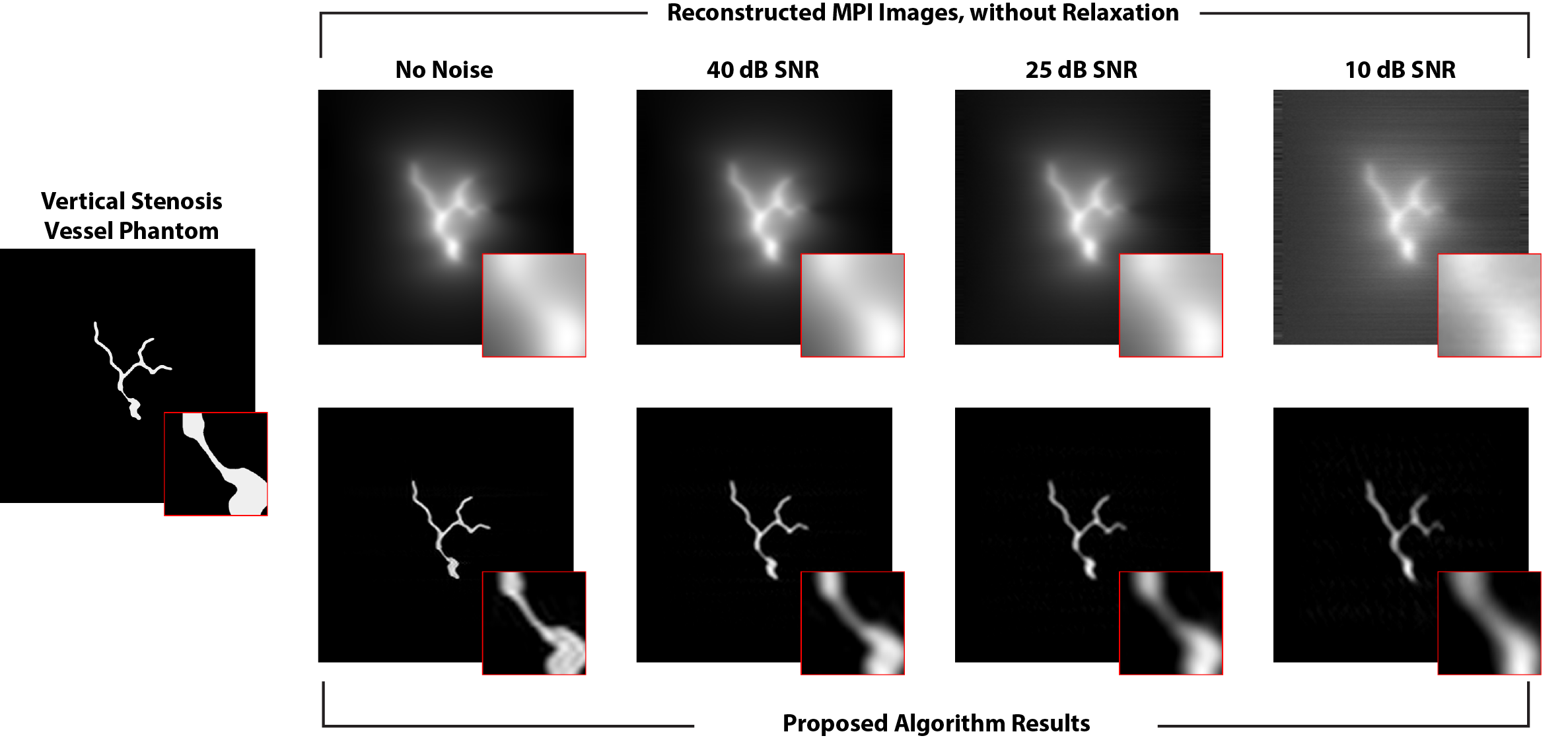}
\caption{Simulation results for MPI images of a vessel phantom with stenosis in a vertically oriented vessel. The vessel phantom (left column). The reconstructed x-space MPI images at varying noise levels, with direct feedthrough filtering (top row). The deblurred images with the proposed blind deconvolution algorithm (bottom row). From left to right: No noise, and 40 dB, 25 dB, and 10 dB SNR. A zoomed-in portion near the stenosis region is shown in small display windows at the corner of each image.\newline
}
\label{results_imgs}
\end{figure}
\begin{figure}[b]
\centering
\includegraphics[width = \columnwidth]{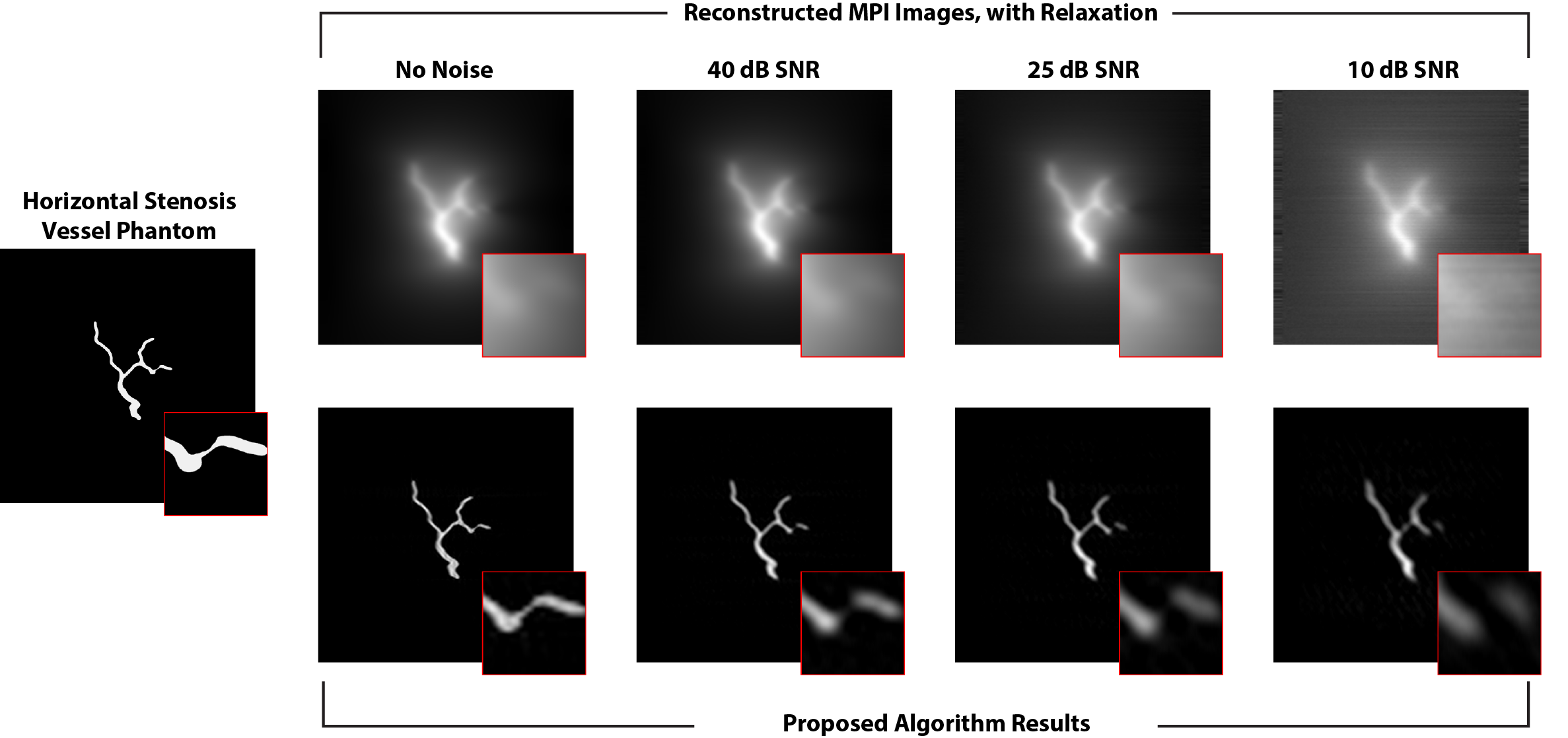}
\caption{Simulation results for MPI images of a vessel phantom with stenosis in a horizontally oriented vessel. The vessel phantom (left column). The reconstructed x-space MPI images at varying noise levels, with direct feedthrough filtering and nanoparticle relaxation effects (top row). The deblurred images with the proposed blind deconvolution algorithm (bottom row). From left to right: No noise, and 40 dB, 25 dB and 10 dB SNR. A zoomed-in portion near the stenosis region is shown in small display windows at the corner of each image.\newline
}
\label{results_imgs2}
\end{figure}

Spatial blurring in MPI images may pose a significant limitation to the accuracy of anatomical assessments. Deconvolution algorithms that recover the original spatial distribution of the nanoparticles from reconstructed images can therefore be critical for many applications including MPI angiography. To examine the success of the proposed deconvolution in vascular assessments, we simulated two phantom images with stenoses in vertical and horizontal directions as shown in Figures \ref{results_imgs} and \ref{results_imgs2}, respectively. MPI images at 9 different SNR levels were reconstructed to test the resilience of the proposed algorithm against noise. For the stenosis in the horizontal direction, nanoparticle relaxation effects were also incorporated.

The two original phantoms, the reconstructed MPI images, and the deconvolution results are shown in Figures \ref{results_imgs} and \ref{results_imgs2}. The proposed algorithm produces deblurred images that accurately depict the vertical stenosis for SNR levels above 10 dB. In the x-space MPI images, especially horizontally-aligned blood vessels are blurred to a wide extent, and the depiction of these vessels are clearly improved in the deconvolved results.
Yet, the horizontal stenosis in Figure \ref{results_imgs2} can only be depicted for SNR levels above 25 dB. This difference is caused by the anisotropic shape of the imaging PSF. As displayed in Fig. \ref{2dRealValued}a, the PSF has much larger spread in vertical direction that severely fades the horizontal stenosis. 

\begin{figure}[t]
\centering
{\includegraphics[width = 0.9\columnwidth]{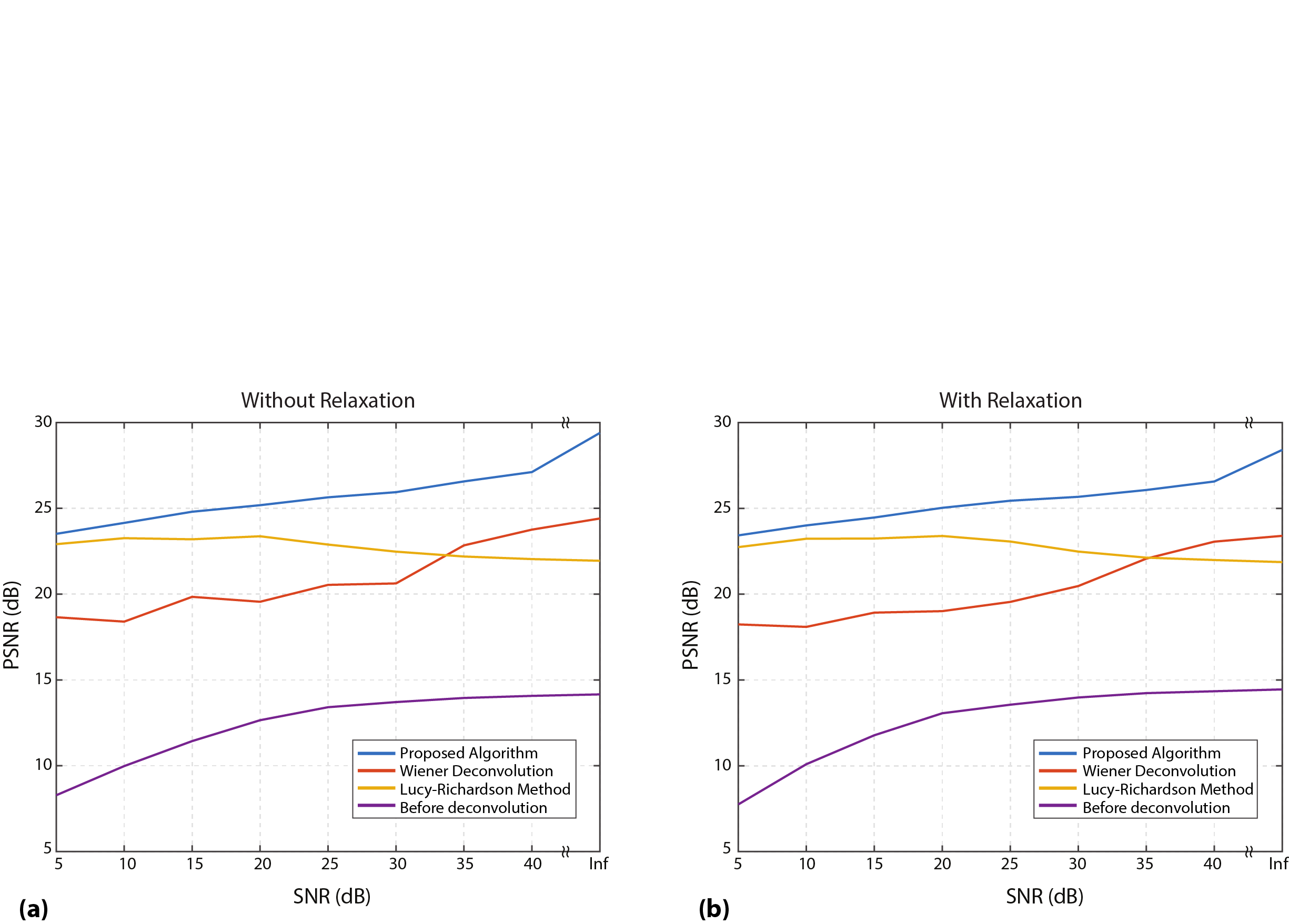}} 
\caption{PSNR as a function of SNR level for the reconstructed MPI images and deconvolution results for (a) the vertical stenosis phantom (without relaxation effects) and (b) the horizontal stenosis phantom (with relaxation effects). On average across SNR levels, the proposed algorithm achieved 5.14 dB higher PSNR than Wiener deconvolution and 2.77 dB higher PSNR than Lucy-Richardson method. The results without and with relaxation effects demonstrate similar performances for all three deconvolution methods, implying that the relaxation effect is sufficiently benign. \newline}
\label{graphic_img_full}
\end{figure}

PSNR measurements of reconstructed and deblurred MPI images over a broad range of SNR levels are plotted in Figure \ref{graphic_img_full}. For the phantom with vertical stenosis (without relaxation effects), PSNR values were 29.39 dB, 27.11 dB, 25.64 dB, and 24.15 dB, for no noise, 40 dB, 25 dB, and 10 dB SNR, respectively.  For the phantom with horizontal stenosis (with relaxation effects), PSNR values were 28.41 dB, 26.57 dB, 25.44 dB, and 24.01 dB, for no noise, 40 dB, 25 dB, and 10 dB SNR, respectively. On average across SNR levels, the proposed method achieved 5.14 dB higher PSNR than Wiener deconvolution and 2.77 dB higher PSNR than Lucy-Richardson method, without utilizing the imaging PSF. Note that the results without and with relaxation effects demonstrate similar performances for all three deconvolution methods, implying that the relaxation effect is sufficiently benign.

\begin{figure}[t]
\centering
{\includegraphics[width = 1\columnwidth]{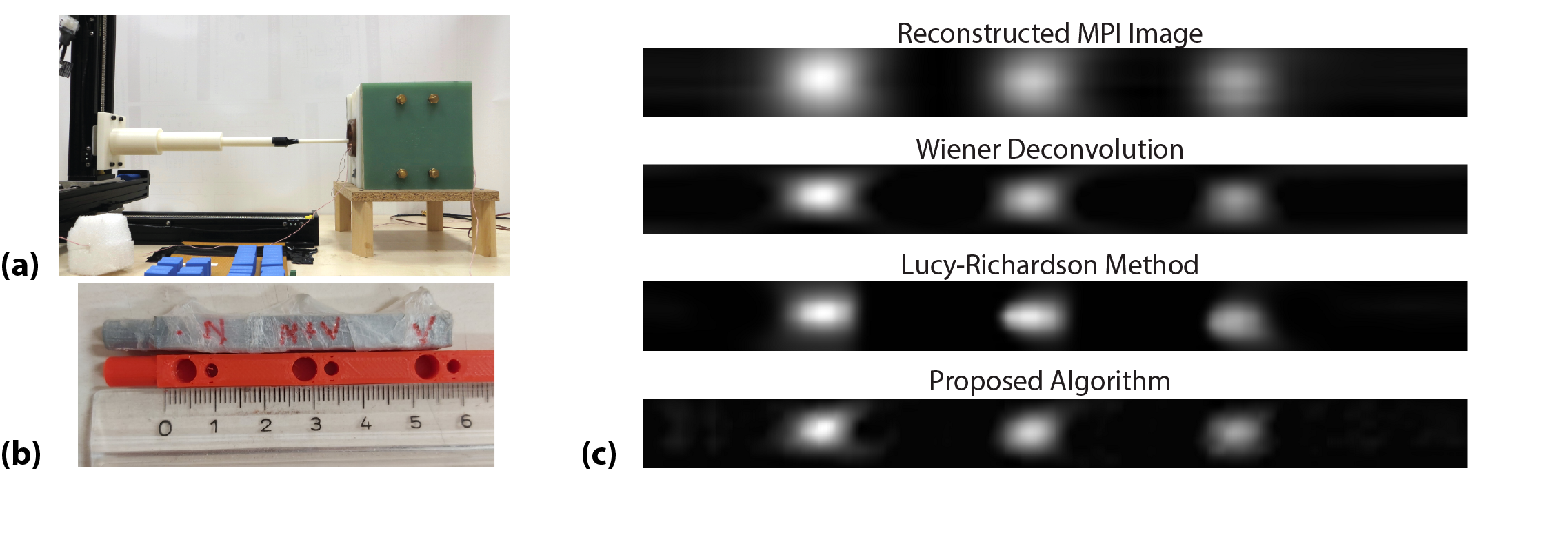}} 
\caption{Imaging experiment results. (a) The experiments were performed on the in-house FFP MPI scanner, using (b) an imaging phantom containing three 2-mm inner diameter vials, placed at 23-mm center-to-center separations. The vials contained Nanomag-MIP (left), Vitotrax (right), and a homogeneous mixture of the two nanoparticles (middle). (c) The reconstructed MPI image, and the results of Wiener deconvolution, Lucy-Richardson method and the proposed algorithm. The FOV was 0.8 cm $\times$ 11 cm in x-z plane.\newline}
\label{graphic_realTest}
\end{figure}

The imaging experiment results are shown in Figure \ref{graphic_realTest}, together with pictures of the in-house FFP MPI scanner and the three-vial imaging phantom used in the experiment. Visual inspection shows that the proposed algorithm yields better delineation of the vials. In these experiments, the performance criteria was chosen to be the full-width-at-half-maximum (FWHM) metric, measured locally for each vial in the images. The FWHM values for the reconstructed MPI image, and the resulting images from the Wiener deconvolution, Lucy-Richardson method, and the proposed algorithm are given in Table \ref{FWHMtable}. The reconstructed MPI image showed FWHM value of 8.09$\pm$0.35 mm (mean $\pm$ STD across vials). The FWHM values after deconvolution were 6.00$\pm$0.31 mm for Wiener deconvolution, 6.12$\pm$0.39 mm for Lucy-Richardson method, and 5.07$\pm$0.09 mm for the proposed algorithm. While the vials originally had 2-mm inner diameter, the presence of noise and relaxation effects cause higher FWHM estimations than ideal. Nonetheless, the proposed algorithm yields the smallest FWHM value, demonstrating its improved resolution capability. These trends validate the simulation results. Furthermore, the FWHM values across vials show higher uniformity for the proposed algorithm, demonstrating its capability to handle different nanoparticle types even in the same MPI image. Note that the proposed algorithm achieves these results without utilizing the imaging PSF.

\begin{table}[]
\centering
\begin{tabular}{l|p{3cm}|p{3cm}|p{3cm}|}
\cline{2-4}
& Nanomag-MIP & Mixture & Vivotrax \\ \hline
\multicolumn{1}{|l|}{Reconstructed MPI Image}  & 8.48 & 8.00 & 7.80\\ 
\multicolumn{1}{|l|}{Wiener Deconvolution} & 6.32 & 5.98 & 5.70\\ 
\multicolumn{1}{|l|}{Lucy-Richardson Method} & 6.36 & 6.33 & 5.67 \\ \multicolumn{1}{|l|}{Proposed Algorithm} & 4.97 & 5.15 & 5.10 \\ \hline
\end{tabular}
 \label{FWHMtable}
 \caption{Comparison of FWHM values (in mm) for the three vials shown in Figure \ref{graphic_realTest} (c).\newline \newline
}
\end{table}

\section{DISCUSSION AND CONCLUSION}
In this study, we showed that the imaging PSF in x-space MPI can be approximated as a zero-phase filter in Fourier domain. We then proposed a new deconvolution technique that leverages this critical phase property. Our technique is based on iterative projections onto multiple convex sets, which reflect various constraints in Fourier and image domains. These constraints enforce consistency of Fourier-domain phase between blurry and deconvolved images, and enforce an estimated spatial-domain support and low $\ell_1$ norm for the deconvolved image.  

The proposed algorithm offers numerous benefits over conventional deconvolution methods such as Wiener or Lucy-Richardson. First, unlike the conventional methods, the proposed algorithm does not require any prior knowledge of the imaging PSF. It thereby performs blind deconvolution of MPI images. Second, it employs the computationally efficient FFT algorithm to transform between image and Fourier domains, and a fast approximate projection onto the $\ell_1$ ball to eliminate the complex sorting of pixel intensities. Thus, it shows relatively improved convergence properties and computational efficiency. Lastly, the spatial regularization based on $\ell_1$ norm in the image domain ensures enhanced reliability against high levels of noise in MPI images. Following similar motivations, $\ell_1$ regularization of MPI images was previously utilized in the solution of the optimization problem posed by SFR \cite{storath2017}, including our own works in \cite{ilbey2017,ilbey2019}. Note, however, that $\ell_1$ regularization for the deconvolution of x-space reconstructed MPI images were previously not shown.

A number of technical advances can help further improve the deconvolution performance. The iterative POCS algorithm is initiated with the reconstructed MPI image. POCS theorem states that it is possible to start from any image and find a possible solution. Note, however, that if projected convex sets intersect at multiple points, the algorithm may converge onto distinct solutions depending on initialization. Here, we observed that initialization with the reconstructed image yields more desirable and consistent solutions than random initialization. It may further improve convergence if the algorithm was instead initiated with the Wiener deconvolution results estimated via an approximate PSF. 

It would also be possible to improve performance if the exact PSF was available to the proposed algorithm. Here, in gradually fading the support during Fourier phase extraction, we used a generic Gaussian PSF. The replacement of this filter with the known PSF may help improve the precision of the estimated phase information, therefore improving the results. Since the algorithm would not be blind in that case, this approach was not investigated in this work.

The proposed algorithm employed $\ell_1$ regularization in the image domain for alleviating noise that corrupts MPI acquisitions. While $\ell_1$ regularization of pixel intensities was observed to perform well here, other types of regularization terms could be incorporated to further improve noise resilience. For example, total-variation terms that assume a block-wise tissue structure can be powerful in noise suppression. Alternatively, $\ell_1$-norm in a transform domain (e.g., wavelet domain) could be used to better enforce sparsity of MPI data. These terms can simply be included in the POCS algorithm as additional projections. 

Due to the relaxation effects in MPI, the transformation between the ideal and blurry images may not strictly satisfy the zero-phase property. Here, the simulation results that incorporated relaxation demonstrated similar performance as those that ignored it, implying that this effect is sufficiently benign. For the cases when the phase of the transfer function is not exactly zero but small in a sufficiently wide region in the central part of the Fourier domain, the zero-phase assumption still leads to reasonable performance for the phase recovery step. For example, a PSF that causes a slight shift in the position of the MPI image would possess such a property. For such a shift scenario, the higher spatial frequencies where the phase of the transfer function is large have corresponding magnitudes that are close to zero. Accordingly, the contribution of such high frequency components to natural images is limited.

Here, we demonstrated a blind deconvolution algorithm for MPI based on simulations incorporating noise, relaxation effects, and complex tissue structure (e.g., stenoses), as well as imaging experiments on a phantom containing different nanoparticle types. Our results strongly indicate that the proposed algorithm can enable accurate mapping of particle distribution in MPI. Numerous MPI applications from angiography to stem cell tracking can then benefit from the enhanced spatial acuity of deblurred MPI images. Further studies should be conducted to validate the current results and assess the reliability of the technique for in vivo imaging. 

\section{Acknowledgments}
The work of E.U. Saritas was supported in part by a Marie Curie Actions Career Integration Grant (PCIG13-GA-2013-618834), by the Turkish Academy of Sciences through TUBA-GEBIP 2015 program, and by the BAGEP award of the Science Academy. The work of T. \c{C}ukur was supported in part by a Marie Curie Actions Career Integration Grant (PCIG13-GA-2013-618101), by a European Molecular Biology Organization Installation Grant (IG 3028), by the Turkish Academy of Sciences through TUBA-GEBIP 2015 program, and by the BAGEP award of the Science Academy.

\section{References}

\bibliographystyle{medphy.bst}
\bibliography{references}

\end{document}